# Evaluation and Performance of Reactive Protocols Using Mobility Model

**Naveed Anjum[1] Imran Shafi[2] and Sohail Abidi[3]**

**[1] Department of Computing and Technology**
**IQRA University, Islamabad, Pakistan**
**naveed.anjum71@yahoo.com**

**[2] Department of Computing and Technology**
**Abasyn University, Islamabad, Pakistan**
**Imran.shafi@abasyn.edu.pk**

**[3] Foundation University Rawalpindi Campus, Pakistan**
**rsohailabid@yahoo.com**

## Abstract

A Mobile Ad-hoc Network (MANET) is a self-motivated wireless network which has no centralized point. It is an independent network that is connected by wireless link so, in which every point or device work as a router. In this network every node forward the packets to the destination as a router and it's not operating as an ending point. In this network every node adjusts them self by on his way in any direction because they are independent and change their position regularly. There are exist three main types of routing protocols which are reactive, proactive and final is hybrid protocols. This whole work compares the performance of some reactive protocols which also known as on - demand protocols, which are DSR, AODV and the final is AOMDV. DSR and AODV are reactive protocols which connected the devices on the network when needed by a doorway. The AOMDV protocol was designed for ad hoc networks whenever any route or link fail and also maintain routes with sequence numbers to avoid looping.

## 1. Introduction

A mobile ad hoc network (MANET) is a grouping of digital nodes that can communicate with each other through wireless transceiver and without using a central point, equipped with wireless transceivers that can communicate with one another node or host without using any fixed networking or central infrastructure [1]. The data packets are sent over wireless Communication control to make communication. A point which makes a big difference and also advantage from other local area networks is that they have no centralized point or base station. Whereas, in cellular networks the communication from a mobile terminal is done by a centralized point and in these networks the mobile terminals make communications directly with each other in the radio transmission range. For transmission to a node that is found out of range of radio transmission, work on the principle of stored and forward mechanism, so these networks also called multi hopes wireless networks because they use multi hop to send data in the networks. With some amount of overhead and bandwidth use, the route creation and repairing is done under some conditions in the networks [2]. The process of finding a route and also repairing it is a main working point in dynamic source routing protocol [3]. The process of finding a complete route to target node and also acknowledgements from a destination in the networks make successful route between sources to destination [4].

### 1.1 Open issues in MANET's

There are still many problems exist in mobile ad hoc networks. As in this routing environment the efficient routing is affected by mobility of network under security and time because all nodes in the networks act like a router and work on the principle of store and forward the packets from other nodes in the networks. There is another issue exist in these networks is the consumption of energy because every node which lies in the networks not only send the data but also send data from other nodes.

### 1.2 'MANET' Networking

Wireless communication is an advance technology in these days and becoming more popular than before and this type of technology exists between wireless communications devices such as laptop computers and mobile nodes in a limited range of networks. There are two main advantages of this technology is that prices are low and data rates are much higher which make this type of technology more interesting and also a reason of growth of this technology. In wireless communication, there are exit two types of approach which makes communication between any nodes which allow existing networks to carry voice and also the data. In wireless ad hoc networks a grouping of nodes work on the principle that there is no centralized communication or there is no central supervision in the networks [2]. This type of networks has many importance in a large number of applications such as personal digital assistant networks, in military application such as tanks, planes and the aircrafts etc. and also in civil application and emergency operations

### 1.3 APPLICABILITY OF MANET'S

We can apply mobile ad hoc network technology in two different areas which are here:

When a new mobile node is added in wired and wireless networks then this technology is used. For example customers in a city who can communicate each other for obtaining rates information's and also a student and employers in universities and company and many other ways exist.

Another area where mobile ad hoc technology is used, where a communication network wanted but due to some reasons it does not exist such as the whole network destroyed due to wars and disasters and some other problems. This technology is used in different field such as in military, police, hospitals and also rescue operation and many other areas. This technology also reduces the cost of networks.

## 2. Background

### 2.1 MANET Routing Protocols

There are three main types of ad-hoc routing protocols, which explain as here.

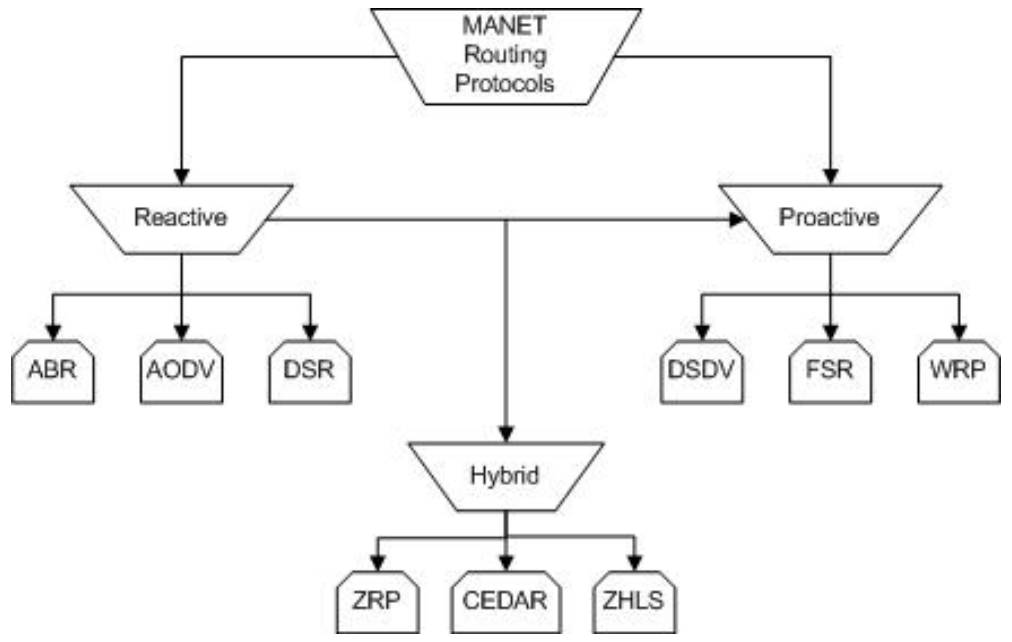

Fig. 1: Mobile ad-hoc routing protocols

#### 2.1.1 Proactive Protocols

The proactive protocols which are also known as table driven protocols work on the bases of fixed mechanism and also used shortest path. In proactive protocol networks, every node lies in the network maintain a routing table which contains all routes to destinations in all over the networks [5, 6]. For updating the table of networks it needs to update the table by sending updating messages to all nodes which lie in the networks. It means the table refresh after a few moments. As, by using routing table which has all routing paths of networks in the table which results consumption of bandwidth and these also have routing overhead which also decrease the bandwidth of networks. So these protocols have a disadvantage which is that bandwidth consumption of networks. There is also an advantage of proactive protocols is that there is always , a route to destination availability of networks. So as a result the delay is very small in the networks.

#### 2.1.2 Reactive Protocols

As, the reactive protocols which are also called on demand protocols whenever, they want to send packet to the destination first they send a request in the whole networks [7]. The characteristics of these protocols are that they find path on demand and exchange information in networks when they required and every route request has replying.

As the route finds to destination on demand so the advantage is that overhead is small which result in low bandwidth use in the networks. There is also a disadvantage of this technique is that when it send a route request in the networks which produce a big delay.

#### 2.1.3 Hybrid Protocols

The hybrid ones are adaptive, and also the combination of two protocols which are reactive and proactive protocols. Reactive protocols are not sensitive to delay and work for network with any movement. While, proactive protocols came with a small delay. From many research it is found that there are still no best routing protocols for all kinds of Mobile ad hoc networks. There are all routing protocols have its own different advantages and capabilities to do work but some specific environments which create a problem because all nodes of networks should be able to work in every environment of the network not for specific. So there are lots of challenges occurs that how can get high performance in every environment. In present many researchers have proposed many hybrid protocols like ZRP, ZHLS and CEDAR etc. [8].

As, I am working on reactive protocols, first one is Ad-hoc on demand vector, and 2nd is Ad hoc on-demand multipath distance vector routing protocols and 3rd is Distance routing protocols which are studied here.

### 2.2 (AODV) Ad-hoc on demand distance vector:

This protocol is a type of reactive routing protocol which work on demand. It is a Single Scope protocol and the working of this protocol is based on *Destination-Sequenced Distance-Vector (*DSDV). The improvement has been done by minimizing the number of source requests which are used to create a route in the networks. Minimizing the number of broadcasts required to create routes in the network. As this is a reactive protocol so that those nodes which are not used in required path they don't participate and also don't maintain route in the network because there are no need of their participation. When a source node or any device want to transmit or send two packets to destination, first it broadcasts a route request in the whole networks then this request goes to its neighbor nodes and same way it moves by different intermediary node to the required destination in networks. As the result every node reply come and then desired path is selected. This process is known as a path discovery process. (Figure 2(a)). It uses sequence numbers to prevent loop just like DSDV.

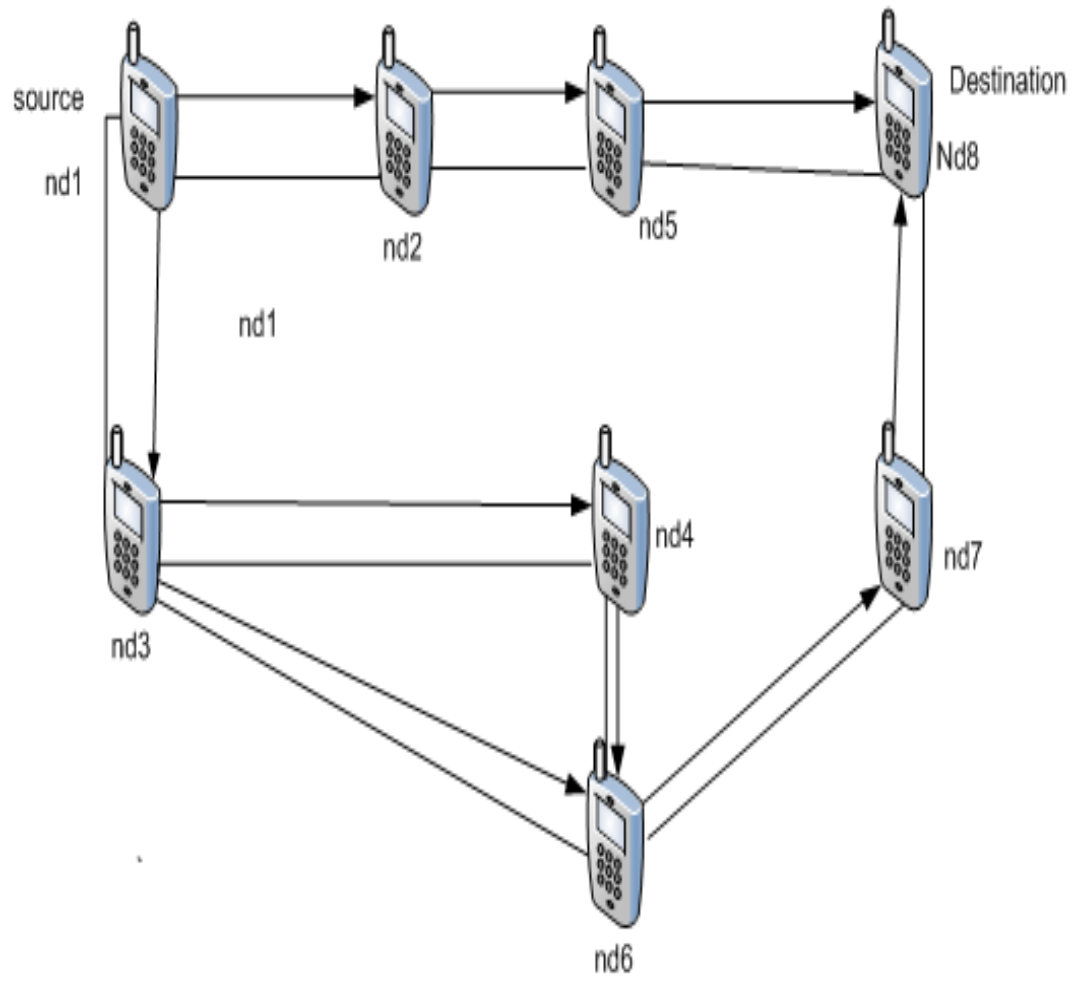

**Figure 2(a) (AODV): How to create a route request [RREQ]**

Each node in the network has two information's, the first information is sequence numbers which prevent from looping and the second is broadcast id which increases a broadcast from starting node. The RREQ which is send by source has the following information's which are Source address of sender, Source sequence number, Broadcast id, Destination address, Destination sequence number, and the number of nodes to the destination. The neighbor's nodes only reply to request only if neighbor's nodes have a way to required destination in the networks with sequence numbers. When a request is sent to the destination, then all the intermediary nodes in the network add the address of its neighbor nodes in the routing table (Figure 2 (b)). If they have equal path then those requests which received later from other neighbors in the network are deleted.

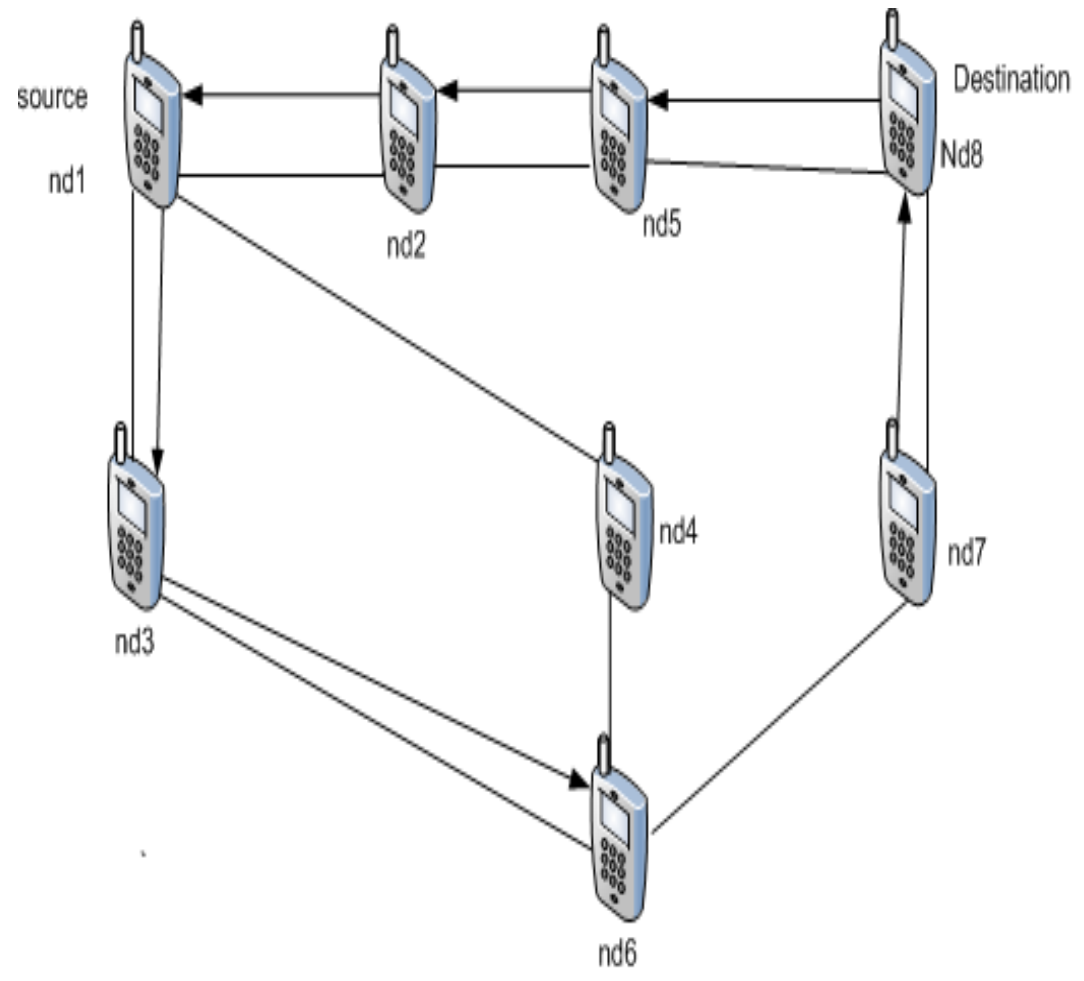

**Figure 2(b):- Reply path to source**

When a source sends request to the destination then all nodes which come between the source and destination, they send a reply to that node from which request come. By this process all midway nodes send a reply to its neighbor from which request come and then final it reached for the source. A route reply that comes from midway nodes have the following information's which are seq no. of the target node, total nodes that come between source to destination, Source address, Target node address, Number of Hops to the destination, Sequence number of the destination and time of expiration for the Reverse Path (Figure 2.8). To establish a link between source and target node, every node which lies between them send a route reply to that node from which request come to establish a path. Therefore this reactive protocol has a bidirectional link which have a path for request send and also route reply. If a broadcast node is missing due to some reasons in the network then route request send again in the networks to maintain the path between source and destination.

**Advantages**

• As there is no updating after some time in the routing table so, in AODV the overload is small in term of packet that's why it is called reactive or on demand protocol. The calculation of this protocol is small because it uses simple messages.
• The purpose of this reactive protocol to provide shortest and fresh path.

**Disadvantages**

• As this reactive protocol establish a path between those nodes who want to establish a path for communication. As this protocol also works on the basis of bidirectional dealing and also uses the route request for sending packets to target node so the disadvantage is that the routing overhead is higher than Dynamic Source Routing protocol.

## 2.3 THE DYNAMIC SOURCE ROUTING (DSR):-

This is another type of reactive routing protocols which using a technique known as source routing. The source nodes that want to send packet to targeted node first find the path of a packet in the network. A packet header has all information's of path from source to destination and also which is going to send packet must determine the path of the packet. The path is attached in the packet header and all information's that are stored in hopes allow for changing with respect to time. When a source node wants to establish a path to another node, it first finds the way to that node on the basis of stored information because there are no updating information's.

**Basic Route Discovery:-**

When a node wants to send a packet to a targeted node it first looks in the route cache which has a previously discovered path. When there are no exist a path in its route cache then the node starts a process for finding the route by sending a route request packet. This packet has the following information's which are, the address of source no which want to send packets and targeted node address and also have a request id. When a node in the network received a route request it first goes into cache to find targets. If there are no exit route from source to target node then this node adds in the route record its address in the route request in the network with the same request id. In Figure 3 (a) how to create a route record when the spreading of route requests the network is shown here.

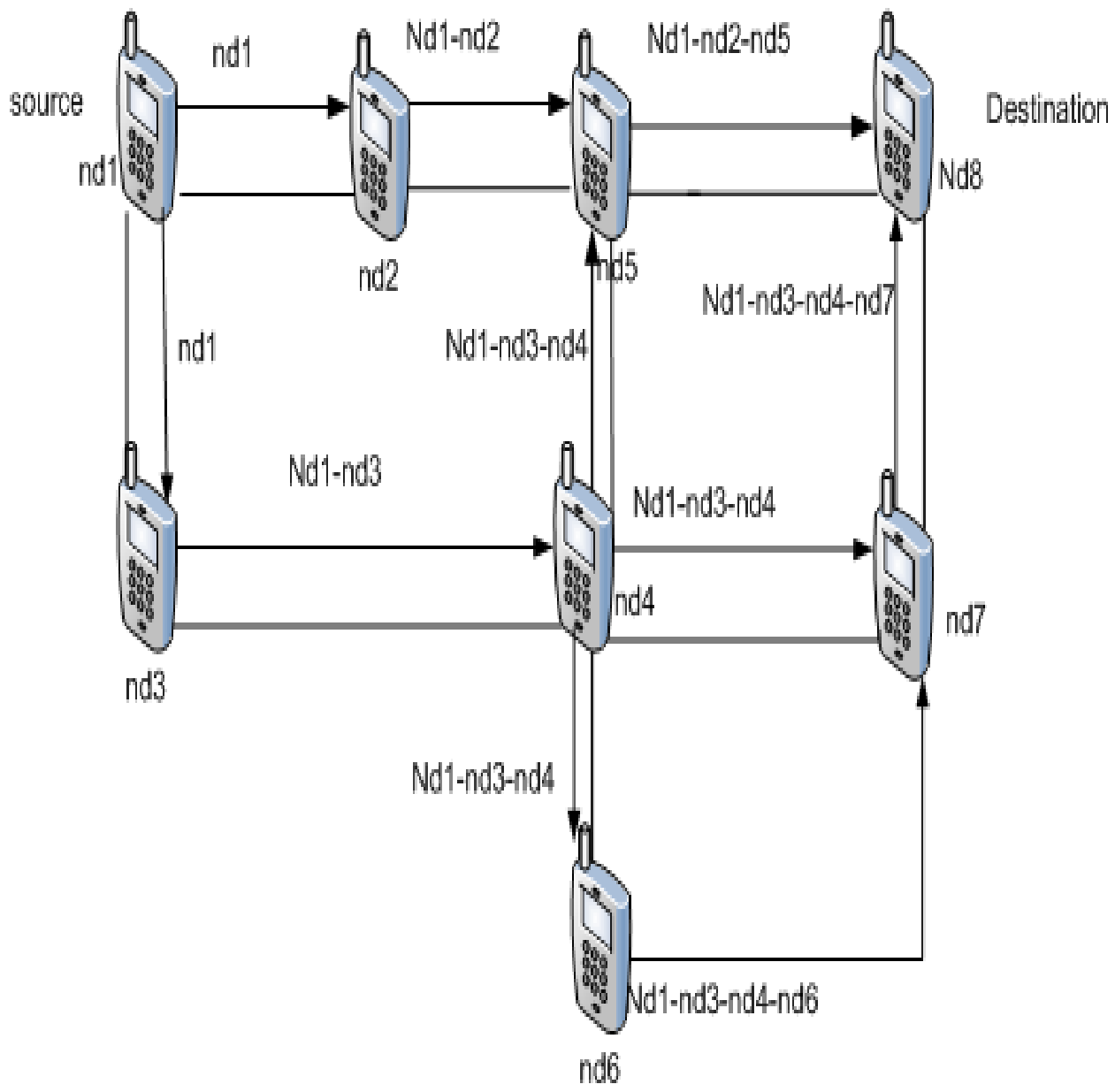

**Figure 3(a):- In route discovery process the creation of the route**

**Record**

Every node attached its address to the route record in the request message and nd1 to nd8 show the address which are added in the route request. When a route request send to targeted node or an intermediate node then in response to this request a route reply come in the network. When a target node has received a route request in a route record and sequence of nodes move in the network. When all nodes to create a route reply to access the targeted node then it copies route record which forwarded at the route request by source shown in Figure 3 (b).

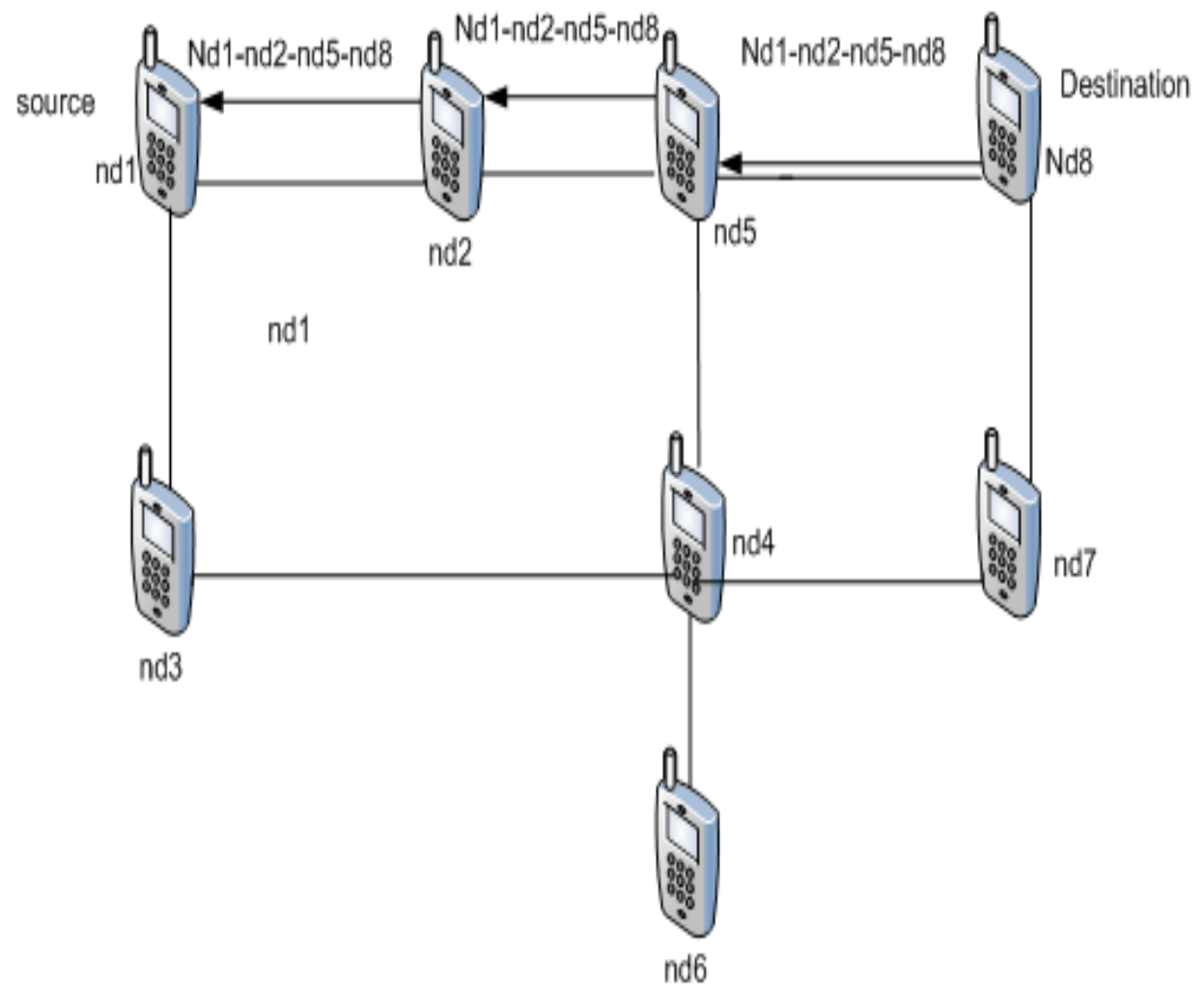

**Figure 3(b):- the movement of route replies with router record.**

### Maintenance of route

The route maintenance is an important factor in the networks. When we find an error in or a link fail on the network then route maintenance play an important role. When there are errors occurs in between nodes of a network then a route error message is sent to the source node. The error message has information's of those nodes in which transmission failed due to link fail. As in figure 3 (c), when nd1 is answerable for receiving of packet at nd2 and also intermediary node nd2 is answerable of receiving packets at node nd5 and then node nd5 is also answerable for receiving of packets at targeted node nd8.

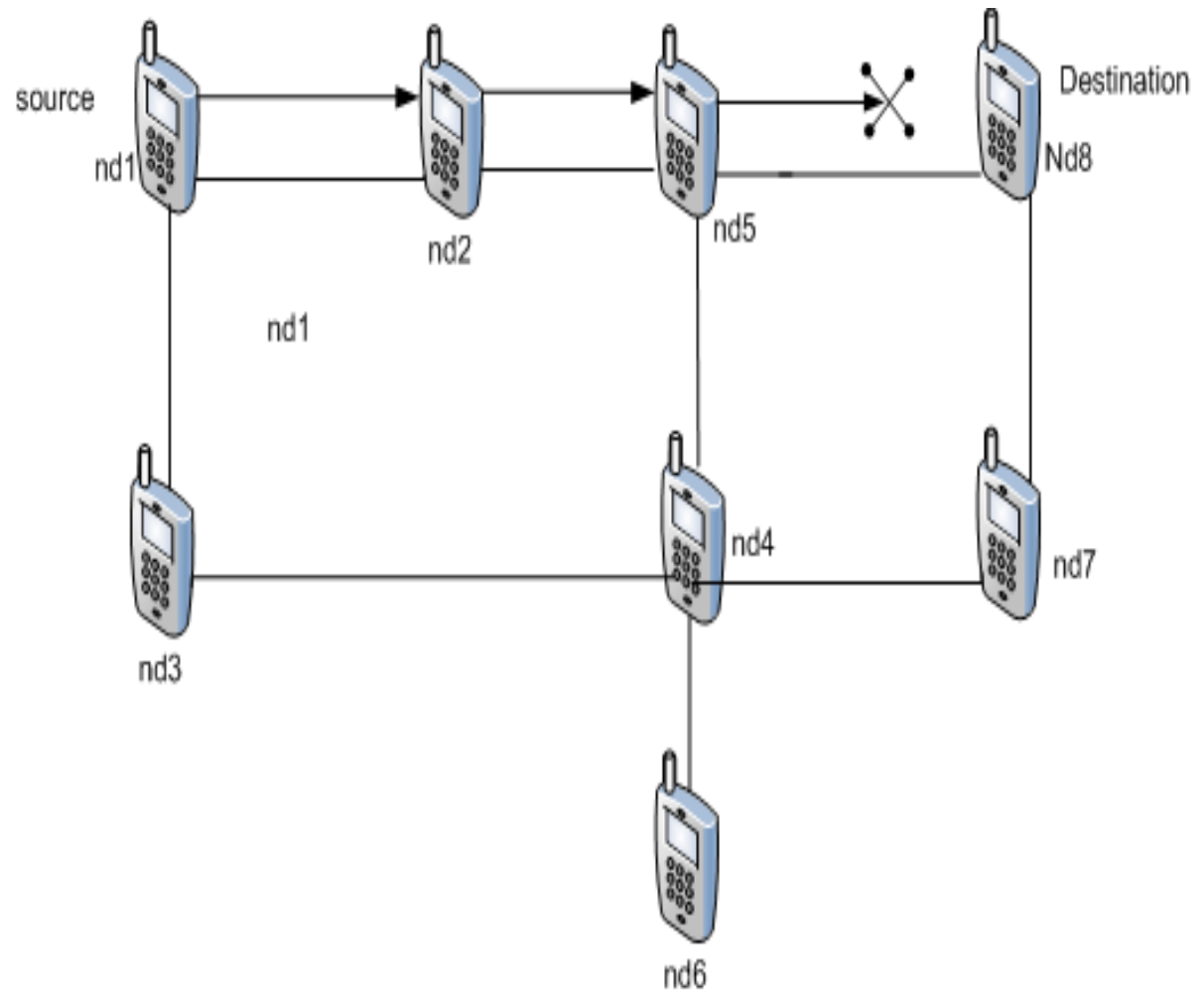

**Figure 3(c):- Route Maintenance exemple:-**

When the intermediate node nd5can not forward packets of the source node nd1 to targeted node nd8 then the intermediary node nd5 send a rout error message to source node nd1 that the link between the nd5 to nd8 is not available or failed due to some reason then the source node nd1 delete this failed link from its cache..

### Advantages

• The 1st advantage is that, when the packets are sent to find route then overload is very minimum because this reactive protocol establish the path only between those nodes which want to make communication.
• 2nd advantage is that only route request process creates routes between source to destination by using cache and intermediary hopes of networking.

**Disadvantages:-**

• 1st disadvantage of this reactive protocol is that when a source node moves a packet to targeted node which use header and header contain the routes. As of result of this if nodes lie in the network is more and more than its result in the form of byte over head.
• 2nd disadvantage is that when a source wants to send the packet's destination, it sends request but this request goes to all nodes lie in the network when this is not necessary.
• 3rd disadvantage is that, this reactive protocol using cache which cause a problem because updating a cache result in overload on the network.

### 2.4 Ad hoc On-demand Multipath Distance Vector Routing (AOMDV):-

This is another type of reactive protocols which is an advanced version of AODV with an advantage of link disjoint path and looping free. It maintains a route to destination on demand. When a route is established to target node then every destination routing entries contain the address record of next node with related hop count As in AOMDV, all subsequent hops in network contain same sequence numbers which is most helpful for continuing the path of route from source to destination. For reaching to the destination every node sends an advertisement in the network to destination and there is a node which makes a hop count. The purpose of hope count for telling that how many numbers hopes come in path to reach the destination. To avoid looping for a node by accepting alternate path to target node if total numbers of hop count less than advertisement hop count. For the same sequence number, the advertisement hop count does not change because this protocol maximum hop count are used to reach to the destination. The advertisement and next hop are restarted when a route advertisement is received with a greater sequence number than the hop count for targeted node. This protocol used to find disjoint route in the network. For finding disjoint route, the every node in the network does not reject duplicate request straight away. A request which come from different destination of network called node disjoint path. As nodes cannot broadcast same request twice, This is because the nodes cannot be broadcast duplicate request, therefore duplicate request which arriving at the midway node by source neighbors cannot traverse the same node. A targeted node answer to duplicate request to get multiple disjoint link and destination only answer to requests which come through by single source nodes.. The advantage of this reactive protocol is that request reply is allowed to midway node while selecting of disjoint path.

Sine, it is multipath routing protocol so its message routing overhead is maximized and also increase due destination replies

## 3. Related work

The Opent, Omenet, GLOMOSIM, NS2 and QualNet are software tools which are used in network base research. S. R. Biradar and his team investigate performance comparison of reactive routing Protocols of MANETs using Group Mobility Model [9]. N.Aschenbruck and his group members perform a survey on mobility models for performance analysis in Tactical Mobile networks in 2008 [10]. The author and his team show a comparison and performance analysis for DSR and AODV in network simulator-2 [11] by using speed, pause time and sources as a changeable parameter. F. Bai and A. Helmy prepared a framework to systematically analyze the impact of mobility on the performance of routing protocols for ad-hoc networks [12]. H. D.Trung and his group performance evaluation and comparison of different ad hoc routing protocols in May 2007 [13]. Yogesh and his team members use GLOMOSIM for comparison and performance analysis of AODV and DSR by using number of nodes, speed and pause time as changeable parameters [14]. Sohail Abid, Imran Shafi and Shahab Khan investigate performance analysis of DSR, AODV and DSDV using RPGM mobility model [15]. Manveen Singh Chadha, Rambir Joon, Sandeep show work in his papers in his paper simulation and comparison of AODV, AOMDV and DSR Routing protocols in MANETs in NS2 [16] by varying from 0 to 50 seconds. , throughput Packet delivery fraction, throughput and the end to end delay are calculated for DSR, AODV and AOMDV. The Harminder S. Bindra1, Sunil K. Maakar and A. L. Sangal , they have done the "Performance Evaluation of Two Reactive Routing Protocols of MANET using the Group Mobility Model" using ns2 by varying speed with respect to the packet delivery fraction, end to end delay, routing overhead, Normalized Routing Load.[17].

## 4. Methodology of Simulation

In the simulation the methodology is used as, The mobility model is used is Random way point mobility model with node maximum speed 0 -60 m/s, the UDP transport protocol is used and simulation time is 300 seconds, and CBR is used as a traffic generator. The simulation area, change in between 500m*500m and node density change from 40 to 80. Node density and simulation area vary from 40 to 80 nodes and 500m x 500m. The other limitations of the simulation show in table 1. For simulation ns2 is used.

| **Parameters** | **Value** |
|---|---|
| Routing Protocols | AODV,DSR,AOMDV |
| MAC Layer | 802.11 |
| Packet Size | 512 bytes |
| Area | 500m * 500m |
| Node s | 50 |
| Mobility Model | Random waypoint model |
| No. of Groups | 4 |
| Data Traffic | CBR, UDP |
| No. of Node | 60, 80 |
| Simulation Time | 300 sec |
| Maximum Speed | 0-60 m/s (interval of 10) |

Table 1

The table contains some parameters in which some are constant and some variable. For testing and verification of result the simulation parameters varying.

### 3.2 Performance Matrices

To evaluate the performance of reactive protocols in my this thesis by using following parameters, which are End to End Delay, Packet Delivery Ratio, Routing Overhead, and Throughput.

### 3.2.1 Packet Delivery Fraction:-

The packet delivery fraction means that the total data packets are delivered to target which are generated by a source. This can be calculated by dividing the number of data packets collect by target which is send by source.

PDF = (Prec/Psnd)*100

Where 'Prec' is total Packets collected & 'Psnd' is the total Packets sent.

### 3.2.2 Routing Overhead:-

It defines the total numbers of control packets that are produced by the routing protocol during the simulation process of networking. Those packets which are sent to network layers is known as routing overhead.

Overhead = number of routing of packets

### 3.2.3 Throughput:-

It means that the total numbers of packets that are recevied divided by number of packets sent during the simulation process of networking.

Thr = Received Packets / Sent Packets

### 3.2.4 Average End-to-End Delay:-

The average end to end delay metric define that a time taken to transmit a data packet in this network from source to target node.

D = ('Tr' –'Ts')
Where 'Tr' is received Time and 'Ts' is sent Time.

## 5. Results

The simulation and performance investigation of DSR, AODV and AOMDV protocols are examined on the basis of following metrics which are: Packet delivery fraction, Routing Overhead, Average end-to-end delay and Normalized Routing Load. The node density is fixed 60 and 80 and node speed varies from 0 to 60 with an interval of 10.

### 4.2 Packet Delivery Fraction:

According to the fig: 4a, it shows that packet delivery fraction is minimum in AOMDV protocol, medium in DSR and maximum in AODV, when the node density is 60 nodes. Whereas, when the node density is 80 nodes, the results are same but the difference between AOMDV and DSR is less.

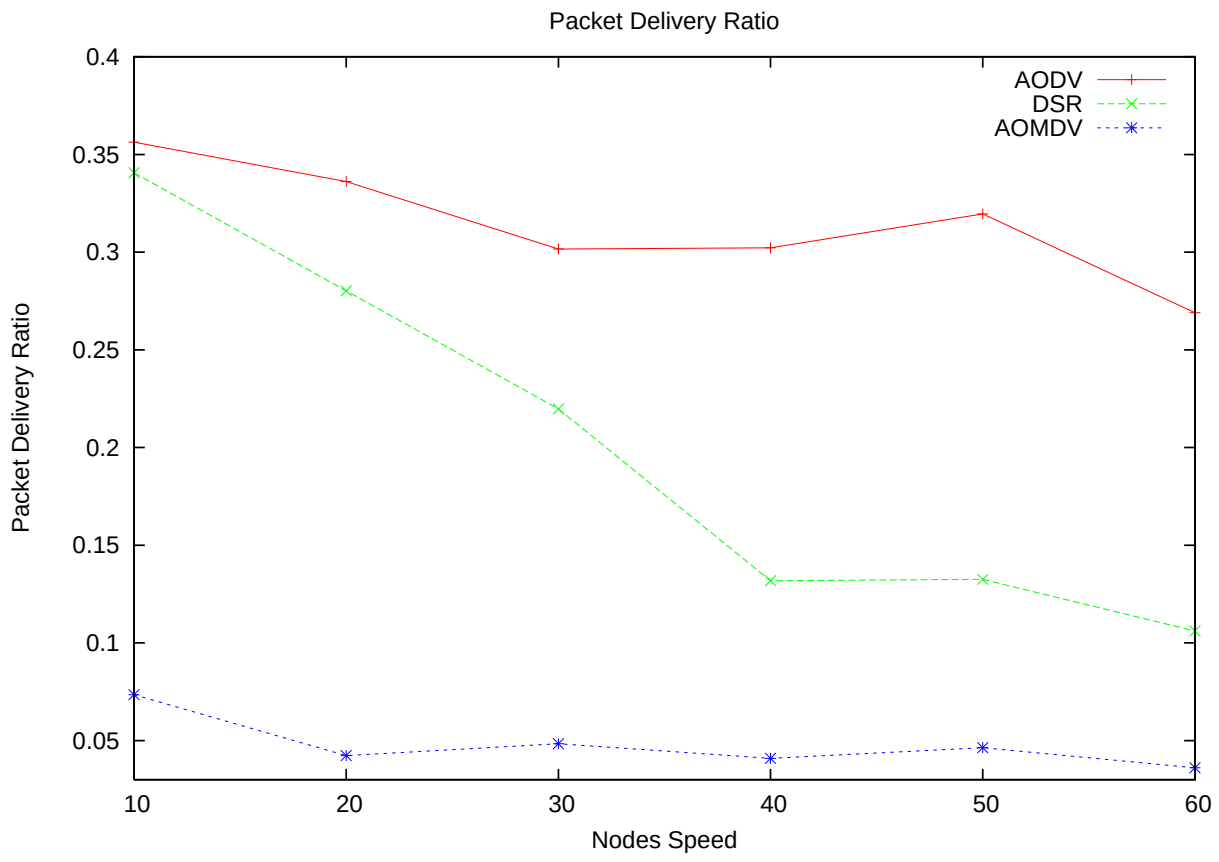

Figure 4a: PDF with 60 nodes

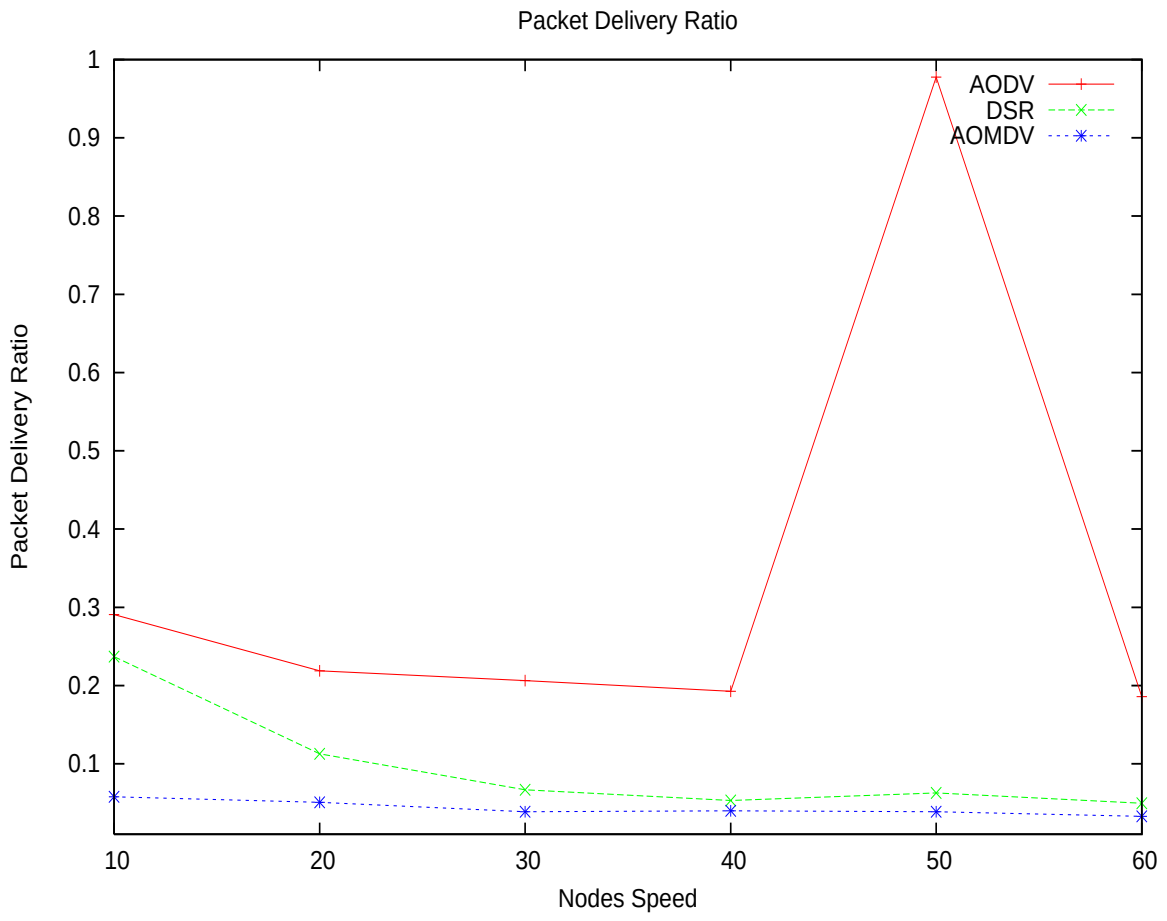

Figure 4b: PDF with 80 nodes

### 4.3 Throughput:

According to the fig: 5a, it shows that packet hroughput is minimum in AOMDV protocol, medium in DSR and maximum in AODV, when the node density is 60 nodes. Whereas, when the node density is 80 nodes, the results are almost same.

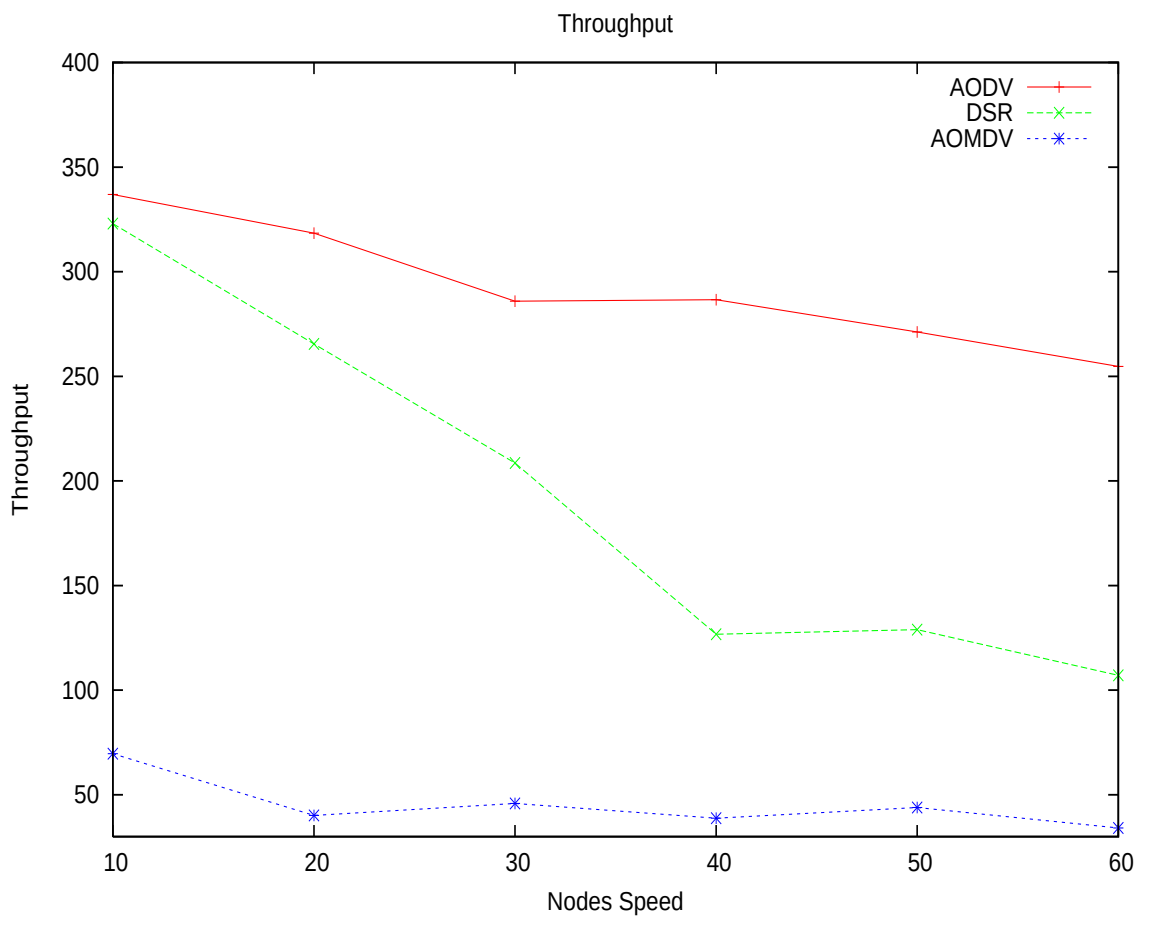

Figure 5a: Throughput with 60 nodes

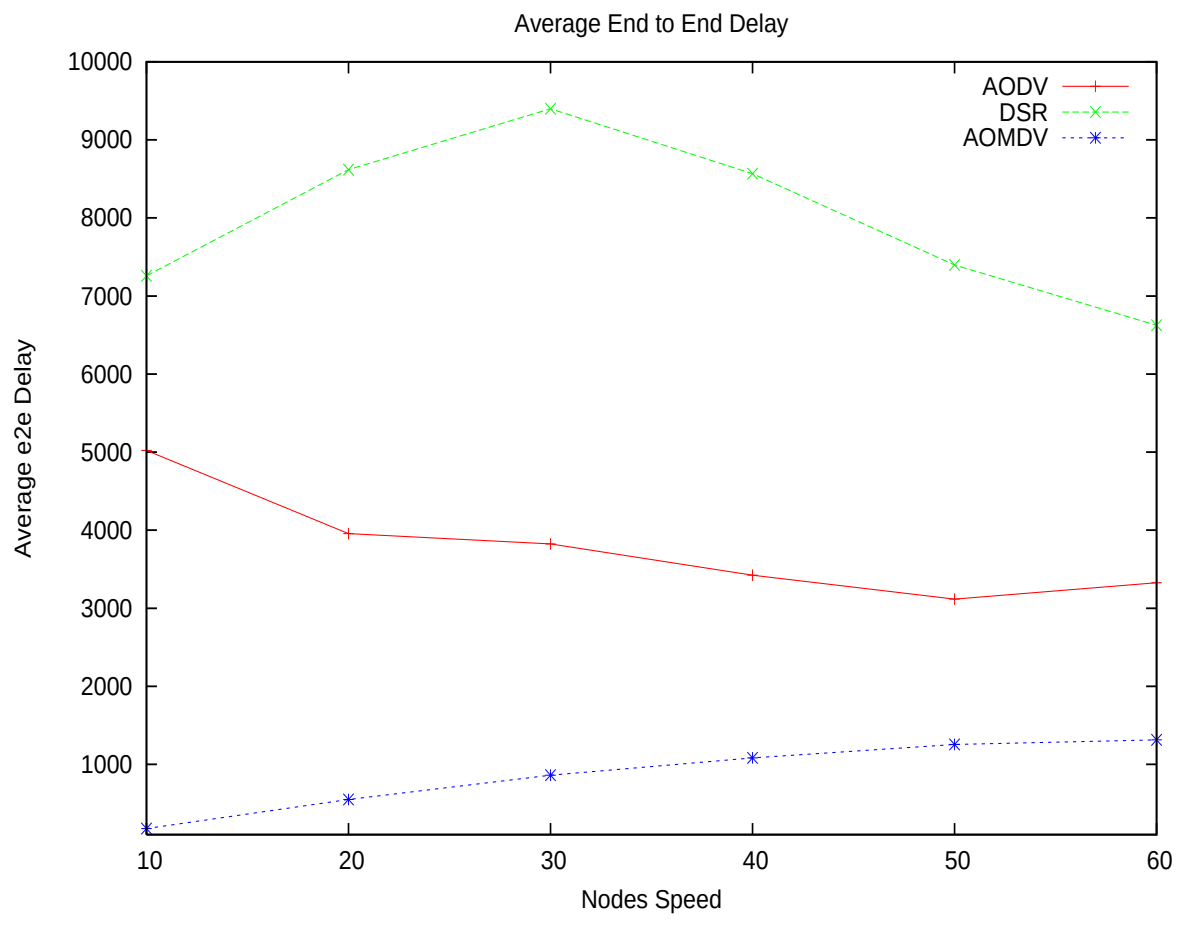

Figure 6a: E2E with 60 nodes

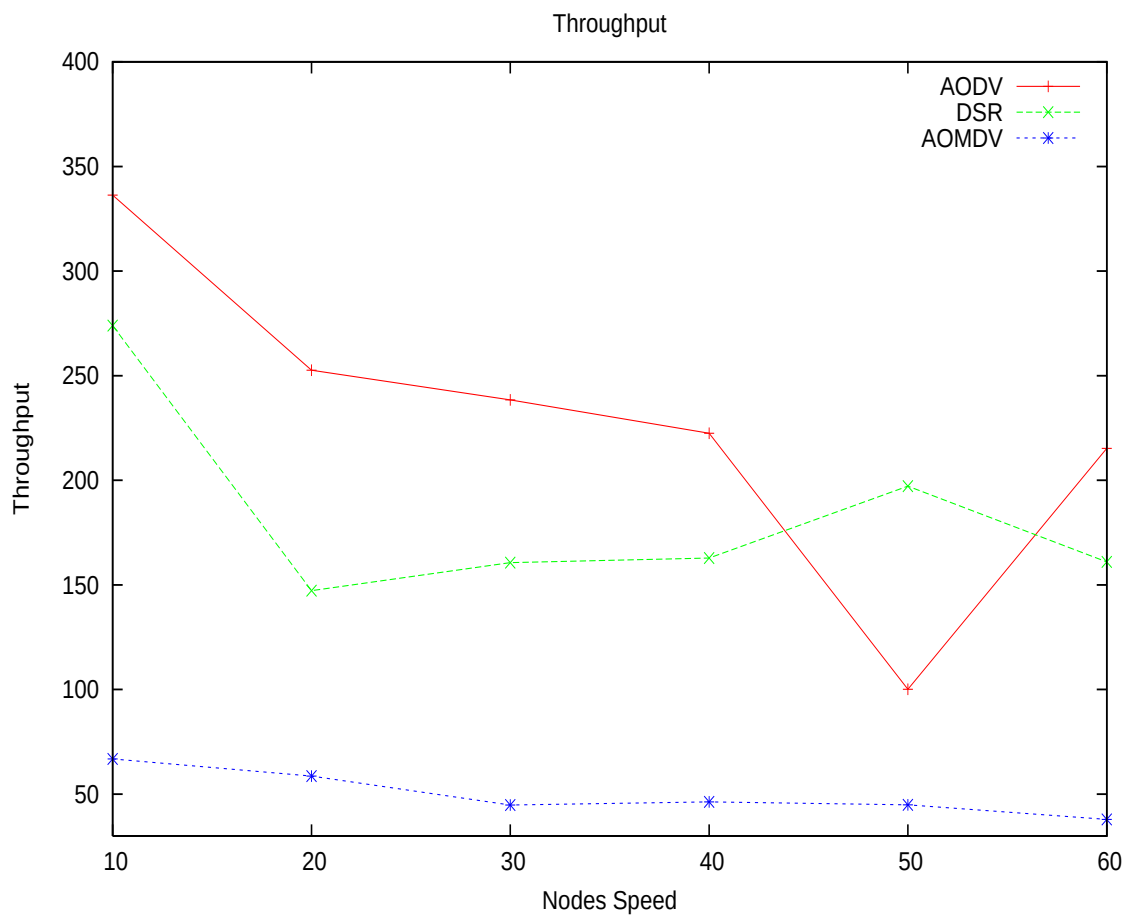

Figure 5b : Throughput with 80 nodes

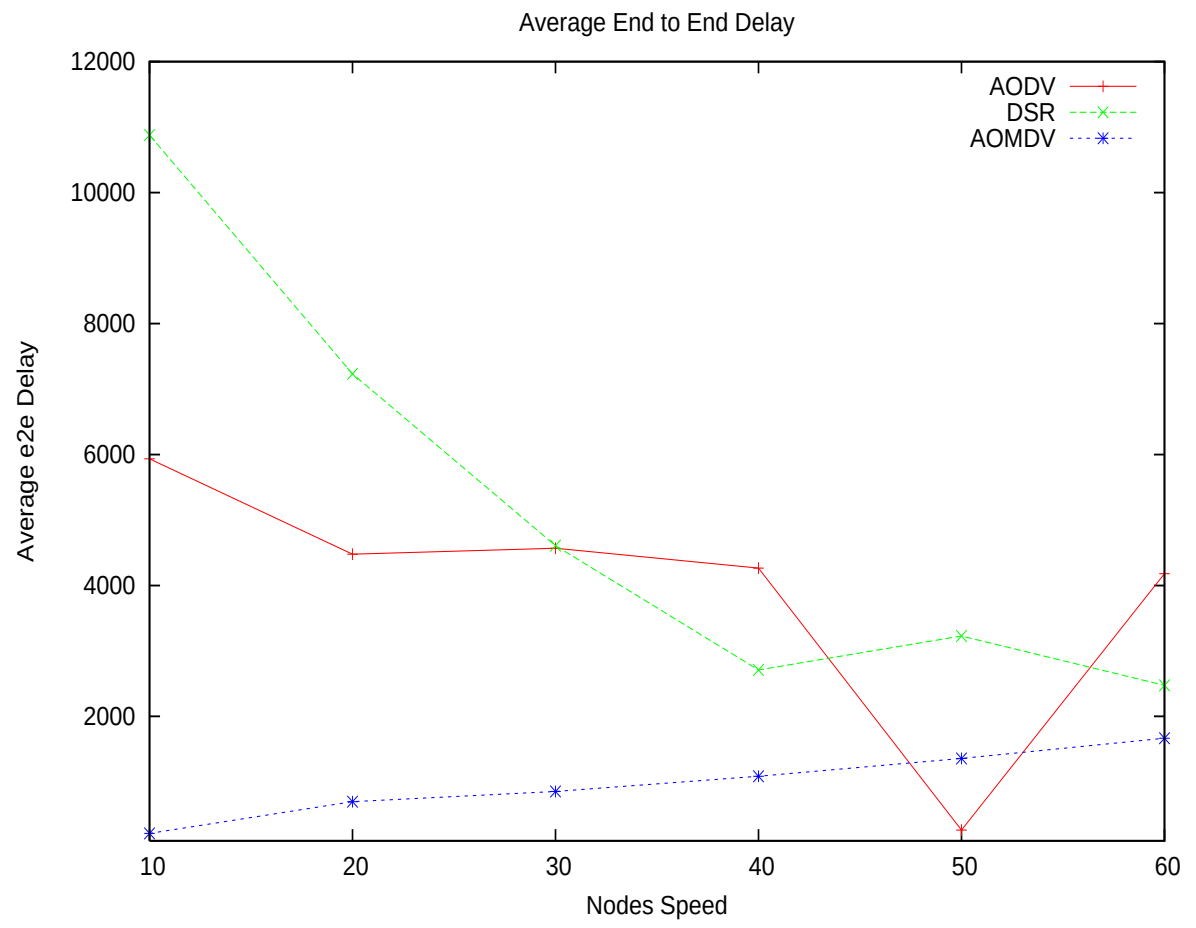

Figure 6b: E2E with 80 nodes

### 4.4 Average End-2-End Delay:

According to the figure 6a when node density is 60, it is experimental proved that Average End-2-End Delay is maximum in DSR, medium in AODV and minimum in AOMDV. But in figure 6b when node density is 80 nodes, AOMDV is minimum, but when node speed is 0 to 30 AODV is medium and when node speed is 30 to 60 DSR is medium.

### 4.5 Routing Overhead:

In figure 7a when node density is 60 nodes, it shows that Rouging overhead is minimum in AODV and medium in DSR. The routing overhead is maximum in AOMDV. In case of AODV routing overhead performance is best.

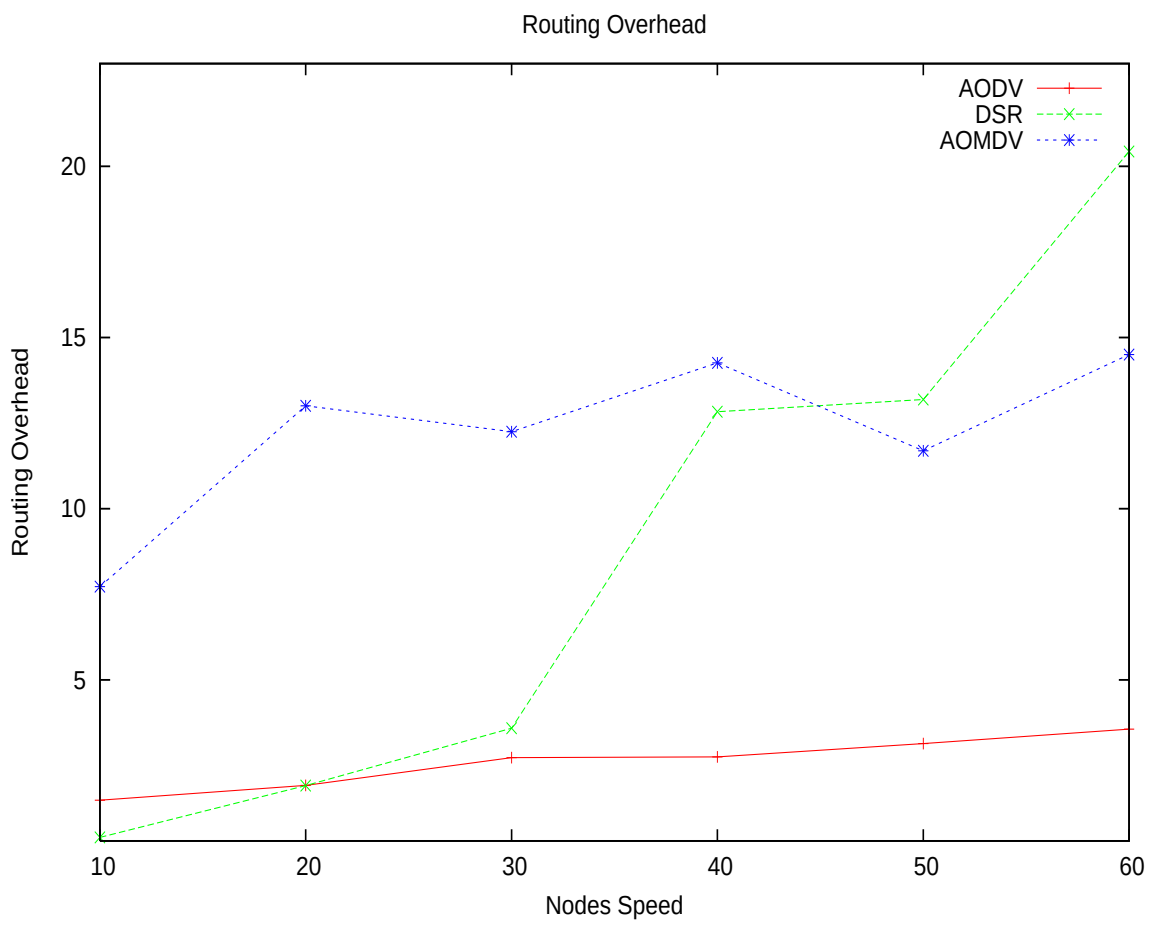

Figure 7a: ROH with 60 nodes

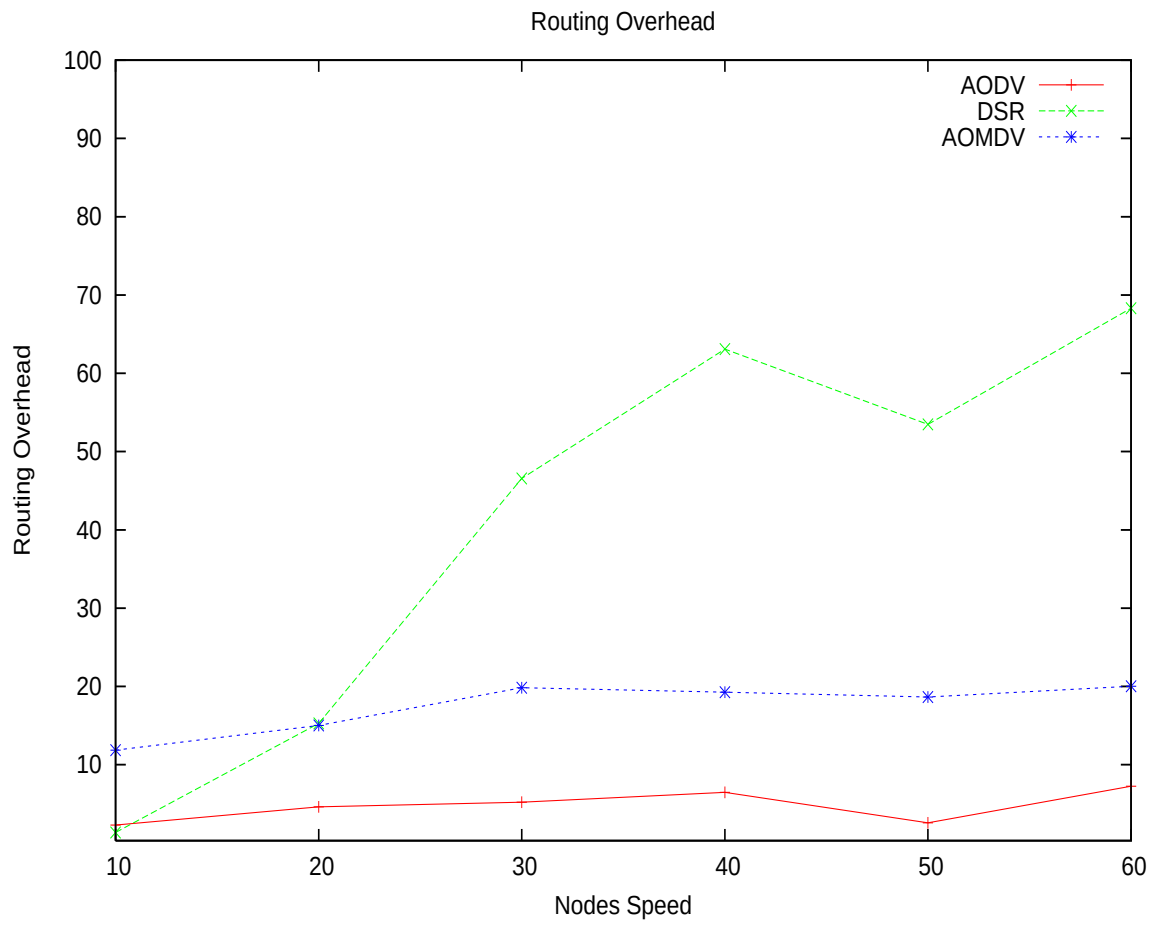

Figure 7b: ROH with 80 nodes

It is interesting to see that performance of "AODV" in routing Through, PDF and overhead is best between the three selected protocols. The "AOMDV" gives best performance in Average end2end delay. The "DSR" gives a comparatively intermediate performance in all simulation results. It is evident from the above figures the performance is not dependent on node density. DSR is the next best protocol and at high node density it is even better than AOMDV. Overall AODV performs better than other protocols in the simulated scenario. The performance analysis of AODV is reasonable in all the four metrics, which makes it clearly the less expensive protocol.

# Conclusion

In this research article we present a summary of MANET protocols and discuss that why performance requires major critical constraints for these types of networks. We also perform a comprehensive investigation of performance analysis metrics and strategies are provided. According to this study it is aiming on four performance analysis methods to attain improved performance. The AOMDV gives best performance in Average end-to-end delay and throughput on the other hand AODV gives best performance in PDF and ROH. The suggestion of this research artical is to build up an proficient performance routing protocol and allows researchers to pick the well describe routing method.

## Acknowledgments

I greatly acknowledged the kind supervision of my teachers and my supervisor, who taught me and encouraged me in this work. I also thanks Mr. Sohail Abid (my co-supervisor).

# References

[1] Pierpaolo Bergamo and Alessandra etl.," Distributed Power Control for Energy Efficient Routing in Ad Hoc Networks ", Wireless Networks 10, 29–42, 2004, Kluwer Academic Publishers, Manufactured in The Netherlands.
[2] N. Pantazis, S. Nikolidakis and D. Vergados, "Energy Efficient Routing Protocols in Wireless Sensor Networks for Health Communication Systems", Ublished in PETRA, pp. 9-13, ACM ISBN: 978-1-60558-409-6, in 2009.
[3] S. A. and P. Tijare, "Performance Comparison of AODV, DSDV, OLSR and DSR Routing Protocols in Mobile Ad Hoc Networks", International journal of information technology and knowledge management, Vol. 2, issue 2, page 545-548, in Dec 2010.
[4] GuntupalliLakshmikanthMr Gaiwak and P. Vyavahare, "Simulation Based Comparative Performance Analysis of Ad-hoc Routing Protocols", INSPEC, ISBN: 978-1-4244-2408-5, page. 1-5, published in 2008.
[5] G. Vijaya Kumar, Y. Vasudeva Reddyr, M. Nagendra, "Current Research Work on Routing Protocols for MANET: A Literature Survey", International Journal on Computer Science and Engineering, Vol. 02, No. 03, pp. 706-713, 2010.
[6] Vincent D. Park, M. Scott Corson, Temporally-Ordered Routing Algorithm (TORA) version 1: functional specification, Internet-Draft, draft-ietf-manet-tora-spec- 00.txt, November 1997.
[7] OLSR, internet draft, http://tools.ietf.org/html/draft-ietfmanet-olsr-00
[8] V. Ramasubramanian, Z. J. Haas, and E. G. Sirer, "SHARP:A Hybrid Adaptive Routing Protocol for Mobile Ad Hoc Networks," The Fourth ACM International Symposium on Mobile Ad Hoc Networking and Computing (MobiHoc), pp. 303-314, 2003.
[9] S. R. Biradar, H. H. D. Sharma, K. Shrama and S. K. Sarkar, "Performance Comparison of Reactive Routing Protocols of

MANETs using Group Mobility Model", International Conference on Signal Processing Systems (IEEE), pp. 192-195, in 2009.
[10] N.Aschenbruck, E.G. Padilla and P. Martini, "A Survey on mobility models for Performance analysis in Tactical Mobile networks," Journal of Telecommunication and Information Technology, Vol.2 pp.54-61, in 2008.
[11] G. Jayakumar and G. Gopinath, "Performance comparison of two on-demand routing protocols for ad-hoc networks based on random way point mobility model," Published in American Journal of Applied Sciences, volume 5, issue no. 6, page no. 649-664, June 2008.
[12] Fan Bai, Ahmed Helmy "A Framework to systematically analyze the Impact of Mobility on Performance of Routing Protocols for Ad hoc Networks", IEEE INFOCOM 2003.
[13] H.D.Trung, W.Benjapolakul, P.M.Duc, ―Performance evaluation and comparison of different ad hoc routing protocols‖, Department of Electrical Engineering, Chulalongkorn University, Bangkok, Thailand, May 2007
[14] Y. Chaba, Y. Singh and M. Joon, "Simulation Based Performance Analysis of On-Demand Routing Protocols in MANETs," Published in Second International Conference on Computer Modeling and Simulation in 2010.
[15] Sohail Abid, Imran Shafi and Shahab Khan, "Investigate Performance Analysis of Routing Protocols Using RPGM Mobility Model", International Journal of Modern Computer Science ISSN: 2320-7868 (Online) Volume No.-1, Issue No.-1, February, 2013. (http://www.resindia.org).
[16] Manveen Singh Chadha, Rambir Joon, Sandeep, "Simulation and Comparison of AODV, DSR and AOMDV Routing Protocols in MANETs", International Journal of Soft Computing and Engineering (IJSCE) ISSN: 2231-2307, Volume-2, Issue-3, July 2012.
[17] Harminder S. Bindra, Sunil K. Maakar and A. L. Sangal, " Performance Evaluation of Two Reactive Routing Protocols of MANET using Group Mobility Model", IJCSI International Journal of Computer Science Issues, Vol. 7, Issue 3, No 10, May 2010.

## AUTHOR'S PROFILES

**Naveed Anjum:** (Mobile No: +92-333-5435768)
Naveed Anjum Student of MS (TN) at IQRA University Islamabad and working as teacher at Aamir Public School.

**Dr. Imran Shafi**: (Mobile No: +92-334-5323402)
Dr. Imran Shafi is working as Assistant Professor at Abasyn University Islamabad.

**Sohail Abid:** (Mobile No: +92-321-5248497)
Sohail Abid is working as System Administrator at Foundation University Rawalpindi Campus.